\newcommand{\buildeb}[1]{\color{RED}{#1}\color{BLACK}}

\newcommand{\loadnormal}{
	\documentclass[prl,twocolumn,epsfig]{revtex4}
	\newcommand{\breakprl}{}
	\renewcommand{\buildeb}[1]{}
}

\newcommand{\loadlengthcount}{
	\documentclass[nofootinbib,prl,twocolumn,epsfig]{revtex4}
	\newcommand{\breakprl}{\newpage \hrule\begin{center} Does not enter the length count:\end{center}\vskip1cm}
	\renewcommand{\maketitle}{}
}

\loadnormal 			% normal version

\usepackage{amssymb,amstext,amscd,amsthm,amsopn,amsmath,mathtools,amsbsy,mathrsfs}
\usepackage{indentfirst,graphicx,epstopdf}

\usepackage[normalem]{ulem} %\sout{Texte � barrer} \xout{Texte � hachurer} \uwave{Texte � ..

\graphicspath{{./figures/}}

\usepackage{color}
 \definecolor{BLACK}{gray}{0}
 \definecolor{WHITE}{gray}{1}
 \definecolor{RED}{rgb}{1,0,0}
 \definecolor{GREEN}{rgb}{0,1,0}
 \definecolor{BLUE}{rgb}{0,0,1}
 \definecolor{CYAN}{cmyk}{1,0,0,0}
 \definecolor{MAGENTA}{cmyk}{0,1,0,0}
 \definecolor{YELLOW}{cmyk}{0,0,1,0}

\newcommand{\Cal}[1]{{\cal #1}}
\newcommand{\be}{\begin{equation}}
\newcommand{\ee}{\end{equation}}
\newcommand{\bea}{\begin{eqnarray}}
\newcommand{\eea}{\end{eqnarray}}
\newcommand{\p}{\partial}

\newcommand{\ind}[1]{{\begin{scriptsize}\mbox{#1}\end{scriptsize}}}

\newcommand{\id}{\mathbb{I}}

\newcommand{\hc}{\mbox{h.c.}}

\newcommand{\ir}{{\textsc{sc}}}
\newcommand{\Hfreeir}{H_0^{\ir}}
\newcommand{\Hbir}{H_{\textsc{b}}^{\ir}}

\newcommand{\ibs}{I_{\textsc{bs}}}
\newcommand{\bs}{{\textsc{bs}}}

\newcommand{\gbs}{G_{\textsc{bs}}}
\newcommand{\TB}{T_{\textsc{b}}}

\newcommand{\HB}{H_{\textsc{b}}}
\newcommand{\NEQ}{{\textsc{n.e}\text{q}}}
\newcommand{\EQ}{{\textsc{e}\text{q}}}
\newcommand{\Ugauge}{\mathscr{U}}

\newcommand{\BJ}{\mbox{B}}

\newcommand{\vph}{^{\vphantom{\dagger}}}
\newcommand{\vf}{v_{\textsc{f}}}

\newcommand{\og}{``}
\newcommand{\fg}{''}
\newcommand{\eq}{\!=\!}
\newcommand{\sh}[1]{\!#1\!}

\newcommand{\eff}{\mbox{\begin{scriptsize}eff\end{scriptsize}}}
\newcommand{\MU}{\Xi}

\newcommand{\UB}{\Ugauge_{\textsc{b}}}

\newcommand{\arxiv}[1]{} %\texttt{arXiv:#1.}}

\newcommand{\density}[3]{: \!#1^\dagger_{#2} #1\vph_{#3} \! : }

\newcommand{\psiir}{\Psi}

\newcommand{\notes}[1]{}
\renewcommand{\notes}[1]{#1}

\newcommand{\XXX}[1]{}
\newcommand{\paragprl}[1]{\emph{#1.}}

\begin{document}

\title{Out-of-equilibrium properties and non-linear effects for interacting quantum impurity systems in their strong coupling regime}

\author{L. Freton}
\affiliation{Laboratoire Mat\'eriaux et Ph\'enom\`enes Quantiques,
Universit\'e Paris  Diderot, CNRS UMR 7162, 75013 Paris, France}
\author{E. Boulat}
\affiliation{Laboratoire Mat\'eriaux et Ph\'enom\`enes Quantiques,
Universit\'e Paris  Diderot, CNRS UMR 7162, 75013 Paris, France}

\begin{abstract}
We provide an exact description of out-of-equilibrium fixed points in quantum impurity models, that is able to treat  time-dependent forcing.
Building on this, we then show that analytical 
out-of-equilibrium results, that exactly treat interactions, can be obtained in interacting quantum impurity models in their strong coupling regime, provided they are integrable \emph{at} equilibrium and they are "super Fermi liquids", i.e. they only allow for integer charge hopping. For such systems we build an out-of-equilibrium strong coupling expansion, akin to a Sommerfeld expansion in interacting systems.
We apply our approach to the Interacting Resonant Level model, and obtain the exact expansion around the low energy fixed point of the universal scaling function for the charge current as a function of voltage, temperature, and frequency, up to order seven.
\end{abstract}

\maketitle

The study of nano structures forced out-of-equilibrium is a vivid field of research, driven amongst other things by long term efforts towards the miniaturization of electronics. Rapid progresses in the realization of engineered  
micrometric structures coupled to macroscopic electrodes (e.g. quantum dots \cite{qdots,hansonreviewQDots}), or of hybrid devices consisting of   atoms or molecules embedded in circuits \cite{atomicelectronics,molecularjunctions,molecularelectronics}
demonstrate the possibility of ``single electron" electronics \cite{kastner92,likharev99}. A theoretical understanding of the mechanisms governing  
transport in those systems is thus of crucial importance. 
Whereas linear response  -- when the voltage across the nanostructure goes to zero --  boils down to equilibrium properties and is fairly well understood, non-linear effects are significantly harder to predict:  
 solving the out-of-equilibrium theory unfortunately turns out to be a considerable problem, mixing  many-body aspects  (interactions make it complicated) with the intrinsically open geometry of the out-of-equilibrium problem. 

\begin{figure}[h]
\includegraphics[width=0.79\linewidth]{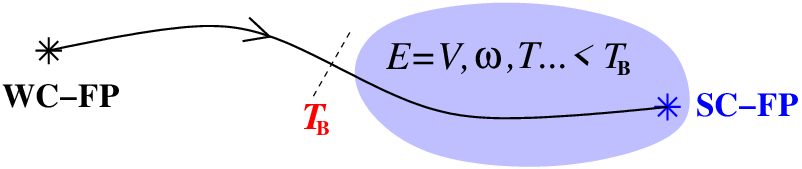}
\caption{A sketch of the renormalization group trajectory that shows a
QIM interpolating between the decoupled (WC-FP) and  strong coupling (SC-FP) fixed points when the energy scale $E$ at which it is probed is varied. Roughly speaking, a point one the curve corresponds to an effective theory with renormalized, effective couplings at energy $E$. The energy scale $\TB$ marks the crossover between the weak coupling regime and the SC regime (our focus, shaded area).}
\label{sketchRG}
\end{figure}

Those systems are modeled by quantum impurity models (QIM), that consist in continua of electronic degrees of freedom representing the metallic electrodes (the baths) interacting with the nanostructure (the ``impurity").  
 A generic feature of QIM at equilibrium is that the impurity/bath coupling, no matter how small, has drastic consequence on the groundstate of the system:  impurity degrees of freedom \emph{hybridize} with the bath, so that effectively some, or all, impurity degrees of freedom are swallowed by the baths. 
  In the zero energy $E\to 0$ (groundstate) limit, impurity degrees of freedom are strongly bound to the baths, so that is is often called a strong coupling (SC)
   fixed point (FP). This last term means that the system has scale invariance in this limit\cite{cardy91,affleck90}, resulting in the fact that the SC-FP is essentially an \emph{homogenous} (the impurity "disappears") and free theory.
 This has direct physical consequences: for example, for systems with a Fermi liquid SC-FP \cite{FLhistorical1,FLhistorical2}, hybridization results in a linear $I(V)$ characteristic at small voltage, with the conductivity $G_0=\frac{\p I}{\p V}\big|_{V\!=\!0}$ being maximal at $T=0$ and at the particle-hole symmetric point. 
 A typical energy scale, called here the hybridization temperature $\TB$ 
 (and akin to a Kondo temperature, the scale below which the impurity spin hybridizes to conduction electrons in the Kondo model \cite{hewson}), marks the crossover between the weak coupling and the SC regime (Fig.\ref{sketchRG}). Technically, the path connecting the high energy (weak coupling) and the SC regime can be apprehended with the renormalization group (RG), and the SC phenomenon reveals itself by a divergence of the effective impurity/bath coupling at low energy\cite{anderson70}.
We run the RG \emph{backwards} and start from the limit $E/\TB\to0$, $E$ being any energy scale at which the system is probed: this is the SC-FP, described by a free theory.  As soon as $E/\TB$ acquires a finite value,  additional many-body scattering mechanisms must be taken into account to describe the physics -- and, incidentally, non-linear effects. 
This SC regime, defined by all energy scales being smaller than $\TB$, is the focus of this work  ;
it is not captured by conventional perturbative expansions, performed around the weak coupling fixed point, that encounter  convergence problems  when the largest of the ratios $E/\TB$ becomes $\lesssim 1$ -- 
 in agreement with the common wisdom that ``the largest energy scale $E$ cuts the RG flow". In this respect, it is important to realize that even a non-interacting resonant level model floes under RG and ends up  entering a SC regime at low energy.

Typically, 
QIM can be brought to effective one-dimensional systems
(by restricting to the $s-$wave scattering or to the relevant conduction channel(s) \cite{affleck90})
homogeneous in space except at one point where the interaction with the impurity is concentrated. This one-dimensional character comes along with a realm of powerful methods   -- be they analytical or numerical -- that gives the hope one could solve the out-of-equilibrium problem. 
Numerical techniques have made significant progresses recently, with the development of efficient time-dependent algorithms. 
Nevertheless, real-time numerical approaches performed on finite systems like time-dependent DMRG face a difficulty in extrapolating to large  sizes or times \cite{irlmNoise}, and methods formulated in infinite systems like diagrammatic  Monte Carlo \cite{diagrammaticMC} or time dependent NRG \cite{TDNRGschiller} run into problems for accessing the stationary state in the SC regime. 

On the analytical side, existing methods like e.g. real-time  RG \cite{korb07,karrash10FRG,schuricht11,hoerig12} or
functional RG \cite{karrash08FRG,jakobs10FRG,karrash10NFRG} either work far from the SC-FP or rely on an approximation to produce explicit results in the SC regime. Exact results, beside their fundamental interest, would therefore be a precious tool to assess the validity of more general, approximate methods.
Since many 
QIM are integrable \cite{bookintegrable} (and thus exactly solvable) at equilibrium, one could hope for a full exact solution  out-of-equilibrium. While this is in fact the case for free (i.e.\ non-interacting) QIM (with a
hamiltonian quadratic in fermions, a very special subset of integrable 
QIM) where the Landauer Buttiker formalism applies\cite{landauer57,buttiker86,imrylandauer99}, and also for some interacting QIM that are amenable to free 
QIM by possibly complicated, non local transformations\cite{SchillerHershfield95,sela09a}, this might well be a property of free fermion systems. Indeed, there are to the best of our knowledge only two examples of \emph{interacting} 
QIM that could be solved exactly out-of-equilibrium: the Boundary Sine Gordon model, solved by algebraic methods \cite{BLZ99} or Thermodynamical Bethe Ansatz (TBA) \cite{FLSprl,FLSprb}, and the Interacting Resonant Level model (IRLM) at its self dual point (SD-IRLM)
using TBA \cite{irlmCurrent,irlmNoise}. 
It is likely that the integrable structure is not preserved when the system is coupled to reservoirs, except in a few exceptional cases. Moreover, the  TBA approach is limited to static forcing, and cannot address e.g. the experimentally relevant AC regime.

In this Letter, we are going to use integrability in a weaker sense, by showing that it can provide an  \emph{analytic}, \emph{universal}, and \emph{systematic} expansion of physical quantities in out-of-equilibrium conditions (including time-dependent forcing) in the SC regime, that can be pushed to arbitrary order in principle. 
Precisely, we  compute the universal scaling functions determining the physical quantities (e.g. $I=V\,f(\frac{V}{\TB},\frac{T}{\TB},\frac{\omega}{\TB},...)$  for the current) in an expansion in $\TB^{-1}$, building a kind of Sommerfeld expansion in interacting systems.
We implement our approach in Fermi liquids, but it goes far beyond the standard Fermi liquid (FL) approach
\cite{FLhistorical1,FLhistorical2}  that only fixes the first correction to the $T=0$ linear regime. In contrast to what happens at higher dimensions, \cite{chubukov04,chubukov06} we find that higher corrections remain analytic.

We proceed  in three steps: First, the forcing out of equilibrium (that can be dynamical) is exactly incorporated in the description of the SC-FP for arbitrary 
QIM:
we  give explicitly the out-of-equilibrium density matrix.

 Second, we focus on integrable 
QIM and carry on a Keldysh expansion in the distance to the SC-FP. In the case of a super Fermi liquid
in which only  transfer of \emph{integer} charges (in however complex processes) are allowed, analytical properties allow for an \emph{exact}  expression  for the  out-of-equilibrium average value of local operators $\hat{\cal A}(x,t)$ in terms of \emph{equilibrium} average value of effective operators in a free theory $\Hfreeir$:
\bea
&\big\langle\hat{\cal A}(x_1,t_1)...\big\rangle_{\NEQ} =\big\langle\hat{\cal A}^{\eff} (x_1,t_1)...\big\rangle_{\Hfreeir}
\label{defOeff}\\
&\hat{\cal A}^{\eff} (x,t)=\Ugauge_{\NEQ}\cdot \UB \cdot \hat{\cal A}(x,t)\nonumber
\eea
where
 the  operators undergo two operations : a dressing $\UB$ by all scattering processes with the impurity, and a gauge transformation $\Ugauge_{\NEQ}$ that translates the out-of-equilibrium forcing. We give in Eqs.(\ref{defUeq},\ref{defUb}) the explicit form of these super-operators.

Third, we obtain an explicit expression for the effective operators (\ref{defOeff}) in a systematic expansion around the SC-FP, that organizes as a series in 
integer powers of $\frac{E}{\TB}$, where $E$ is any energy scale in the problem: voltage, frequency, temperature, particle-hole symmetry breaking...
In full generality, the value of the radius of convergence of such an expansion is an open question.  In the case of the IRLM investigated below, the radius is finite when there are no interactions, 
even in the case of harmonic forcing \cite{buttiker93,jauho94}, and also 
at the self dual point  in the static limit \cite{irlmCurrent}, so that it gives good confidence that it remains finite for any value of the interaction at least for harmonic forcing.
It might be 0 in other situations: our expansion would then be an asymptotic series.

We shall illustrate this method with the concrete example of the IRLM, for which we carried out this expansion at order $(\frac{E}{\TB})^{-7}$ summing up $\sim 1500$ diagrams using a Mathematica code.
This model\cite{irlm_histo}  has recently earned the status of benchmark in out-of-equilibrium physics in 
QIM\cite{doyon07,irlmCurrent,irlmNoise,irlmNumerics,boulat08} as one of the simplest 
QIM allowing for non equilibrium in the presence of interactions. The IRLM consists of two baths of free spinless electrons coupled via tunnel hopping to a single level, that interacts capacitively with the electrodes via a short range potential with strength $U$. 
After the standard steps of linearizing around the Fermi points and unfolding \cite{refunfolding}, 
the two semi-infinite wires are described by two right-moving free fermionic fields $\psi_{1(2)}(x,t)$ coupled at $x=0$ to the impurity level (with creation operator $d^\dagger$):
\bea
H&=&\sum_{a} H_0[\psi_a]+\HB,
\hspace{0.2cm}
H_0[\psi]=-i\vf \!\!\int_{-\infty}^\infty \!\!\!dx\,\psi^\dagger\p_x\psi^{\vphantom\dagger}
\label{defHBRILM} 
\\
\HB&=&\gamma_a\psi^\dagger_a(0)d +\hc + U \!:\!\psi^\dagger_a\psi_a^{\vphantom\dagger}\!\!:\!(0)\big(d^\dagger d -\textstyle{\frac{1}{2}}\big)+ \epsilon_d \,d^\dagger d,
\nonumber
\eea
A sum over $a=1,2$ is implied in $\HB$, and in the following we set the Fermi velocity $\vf\sh{=}1$ as well as $e\sh{=}\hbar\sh{=}k_B\sh{=}1$.
For $U\eq 0$, one recovers a  free theory, the Resonant Level Model (RLM). In the interacting model, standard manipulations (see e.g. Ref.\cite{boulat08}) show that $U\neq 0$ gives the tunneling term an anomalous scaling dimension $D(U)$ \cite{supplement}.
The free theory (RLM) corresponds to $D=\frac{1}{2}$, whereas $D=\frac{1}{4}$ corresponds to the self-dual point (SD-IRLM). In the following we restrict our attention to the interesting situation $D<1$ where the tunneling term is relevant ; in this case the tunneling amplitude $\gamma$  (one sets $\gamma_1\!+\!i\gamma_2=\gamma e^{i\theta/2}$)  flows  to SC under the renormalization group and reaches the value 1 at an energy scale
$\TB/W\sim (\gamma/\sqrt{W})^{1/(1-D)}$ (with $W$
the bandwidth) : below this scale, the  system enters the SC  regime  (Fig.\ref{sketchRG}). $\TB$ is a non-universal quantity, that we fix via the equilibrium charge susceptibility, $\chi_d\vph=\frac{\p\langle d^\dagger\! d\rangle}{\p \epsilon_d\vph}\big|_{\epsilon_d\vph=0}=\frac{1}{2\pi D \TB}$.
The physics in the SC regime is governed by the SC-FP, that is reached at vanishing energy $\frac{E}{\TB}\to 0$, or by formally setting $\TB= \infty$ (in particular this implies $\frac{\epsilon_d}{ \TB}=0$: the SC-FP is particle-hole symmetric). 

%%%%%%%%%%%%%%%%%%%%%%%%%%%%%%%%%%%%%%%%%%%%
\paragprl{(i) Out-of-equilibrium SC fixed point}
The SC-FP has conformal invariance \cite{cardy91,affleck90} implying it can be described by  \emph{transparent} fields $\psiir_a$, that are free fields of an homogeneous system (physically, homogeneity reflects the wires' perfect hybridization).
The effect of the impurity is encoded in the relationship between 
densities of transparent  ($\mathcal Q_{ab}\equiv\density{\psiir}{a}{b}$) and of original ( $ Q_{ab}\equiv\density{\psi}{a}{b}$) fields:
in the incoming region $x<0$, $\mathcal Q_{ab}(x,t)=Q_{ab}(x,t)$, whereas
 for outgoing fields $x>0$, $\mathcal Q_{ab} (x,t)= \BJ_{ab}^{cd}  Q_{cd}(x,t)$ \cite{supplement}.

We force the system out-of-equilibrium by considering a bath with thermodynamical variables $(\mu_a,T_a)$ that 
specifies the density matrix for incoming states in wire $a$.
Since the system is homogeneous for transparent fields, we readily deduce the density matrix at the  SC-FP \emph{out of equilibrium} 
$\rho^{\ir}_{\NEQ}=\rho_{\NEQ}^{\ir,1}\otimes\rho_{\NEQ}^{\ir,2} $:
\be
\rho_{\NEQ}^{\ir,a}=\exp 
-\textstyle\textstyle \frac{1}{T_a}\{H_0 [\psiir_a]- \int_{-\infty}^\infty dx\;\mu_a \mathcal Q_{aa}\}
\label{rhoIR}
\ee
The seemingly simple form of (\ref{rhoIR}) becomes a complicated, non-local object that mixes the wires once reexpressed in terms of the physical fields $\psi_a$ :  Hershfield's operator \cite{hershfieldsoperator} is not local. 
When $\theta=\pi/2$, one recovers the perfectly transmitting situation $\Psi_{1/2}(x>0)=\psi_{2/1}(x>0)$, and our Eq.(\ref{rhoIR}) coincides with the density matrix derived in \cite{bernarddoyon12,bernarddoyon13} (via the identification $a=1,2=+,-$). We generalize this approach for time-dependent forcing $\mu_a(t)$.
As far as properties close to the impurity are concerned, one can use a trick and equivalently 
couple the system to a space-varying chemical potential 
by replacing $\mu_a(t)\to\mu_a(t-x)$ in (\ref{rhoIR}).
%\XXX{The spatial dependence we introduce by this trick is generically extremely weak in the SC regime $\omega\lesssim \TB$, with e.g. a wavelength $\gtrsim$ 1 m for $\TB \lesssim 100$K.}
Those \og right-moving\fg potentials can be absorbed
by a gauge transformation $\psiir_a(z)\to \Ugauge_{\NEQ}^{-1}(z)\cdot \psiir_a(z)$
that maps  the out-of-equilibrium theory onto a product of decoupled equilibrium ones in the transparent basis.
For a generic operator  
one has $\langle\mathcal A[\{\psiir_a\}]\rangle_{\NEQ}=\langle\mathcal A[\{\Ugauge_{\NEQ}\cdot\psiir_a\}]\rangle_{\EQ}$, where  $\langle\,\cdot\,\rangle_{\EQ}$ is the (product of) Gibbs states with density matrix  $\rho_\EQ^\ir=\bigotimes_a e^{-\frac{H_0[\psiir_a]}{T_a}}$. 
%The gauge transformation 
$\Ugauge_{\NEQ}$ can be written as:
\be
\Ugauge_{\NEQ}(z)=
\Cal R e^{-i\!\sum_a\oint_{\! _z}\! \!  dw \,\MU_a(w) \;\mathcal Q_{aa} (w)}
\label{defUeq}
\ee
where 
$\Cal R e^{\oint_{\! _z}\! \!   d\omega  X(\omega)}=1$$+$$\sum_{k>0}\frac{1}{k!}$$\oint_{\Cal C_1}\! \! dz_1$...$\oint_{\Cal C_k}\! \! dz_k$$\, X(z_1)$$...$ $X(z_k)$ 
is a radially ordered  exponential with contours $\Cal C_k$: 
$|z_1\!\!-\!z|\!>\!...\!>\!|z_k\!\!-\!z|\!>\!0$, and  $\MU_a(x,t)=i\int_0^{t-x}dt'\,\mu_a(t')$ is continued to the complex plane $z=i(t-x)$.

Eq.\ (\ref{defUeq}) shows that forcing the SC-FP out of equilibrium  amounts to a deformation of the conformal field theory (CFT) generated by the U(1) charges $\mathcal Q_{aa}$, just as finite temperature effects are generated by the energy-momentum tensor.
Eqs. (\ref{rhoIR},\ref{defUeq}) straightforwardly generalize to arbitrary SC-FP with charges $\mathcal Q_{ab}$  conserved in the bulk -- including non Fermi liquid fixed points. At the SC-FP, this gives a precise answer to the sometimes raised question "to what extent do voltage and temperature play similar roles" in 
QIM
\cite{TvsV}: they correspond to different deformations of the same CFT describing the SC-FP. 

On fermions, $\Ugauge_{\NEQ}$ generates a phase 
$\Ugauge_{\NEQ}^{-1}(z)\cdot \psiir_a(z)=e^{\MU_a(z) }\psiir_a(z)$. Starting from the current operator on wire $a$ close to the impurity,
$\hat I_a=Q_{aa}(0^-)\;-Q_{aa}(0^+)$,
and acting on it with $\Ugauge_{\NEQ}$, one recovers
for the IRLM \cite{supplement} the linear regime 
$I_a^\ir\sh{=}\langle\hat I_a\rangle_{\NEQ}\sh{=}G_0 \, V(t)$ with $G_0=\sin^2\theta \,\frac{e^2}{h}$ and $V\sh{=}\mu_1-\mu_2$ the voltage.

%%%%%%%%%%%%%%%%%%%%%%%%%%%%%%%%%%%%%%%%%%%%

\paragprl{(ii) Around the SC fixed point}
 We now  consider finite values of $\frac{\mu_a}{\TB},\frac{T_a}{\TB},\frac{\epsilon_d}{\TB}$, that drive the IRLM away from the SC-FP. Inhomogeneity reappears: transparent fields $\psiir_a$ undergo scattering at $x=0$, and we should consider all irrelevant processes allowed by symmetries.
The IRLM is a Fermi liquid, so the lowest order process is an energy-momentum tensor, call it $\Cal O_2$, with coupling $\propto \frac{1}{\TB}$. It determines the back-scattered current $\ibs=I_a^\ir-\langle\hat I_a\rangle\propto V^3/\TB^2$ at lowest order: this is Fermi Liquid Theory. At higher orders $\TB^{-n}$ (e.g. to access the leading temperature dependence $\propto T^2V^3/\TB^4$ of $\ibs$), $\Cal O_2$ is not sufficient and new independent, higher order processes must be considered. 

In general one faces here several difficulties, that prevent any practical calculations: the number of independent processes (and corresponding couplings) grows very rapidly with $n$ ; moreover there is \emph{no} way to fix the couplings -- except
that of the first operator, $\Cal O_2$, since it effectively amounts to a definition of $\TB$.
On the contrary, the IRLM is integrable, and 
its infinity of conserved quantities  $\Cal O_{2n}$ highly 
constraints the allowed processes. It  has been shown  \cite{lesage99a,lesage99b} that a  dual description of the Hamiltonian can be derived:
$H=\Hfreeir +\HB^\ir$ with $\HB^\ir=\frac{1}{2\pi}\sum_{n=1}^\infty\frac{g_{2n}}{ \TB^{2n-1}}\Cal O_{2n}(x=0)$
where the couplings  $g_{2n}$ are known. 
The operators $\mathcal O_{2n}$ are the only independent allowed processes, and  transfer \emph{integer} charges across the impurity: we call such a system a ``super Fermi liquid"
which is a sufficient condition for our expansion.

The dual Hamiltonian has been used to describes equilibrium properties \cite{boulat08,freton13}. We  bring this dual description out-of-equilibrium  and we obtain a systematic expansion for arbitrary correlators $\Cal G(t_1,...,t_N)$ of local operators (expressible in terms of the  charges $Q_{ab}$). To do so, we evaluate $\Cal G$ within  a Keldysh expansion in powers of $\TB^{-1}$, starting at time $t=-\infty$ at the \emph{out-of-equilibrium} SC-FP $\TB=\infty$ and adiabatically  turning on a  finite $\TB^{-1}$ value. 
Using a super-operator formulation \cite{harbola08}, we can show \cite{supplement}  that in a super Fermi liquid the full Keldysh expansion can be implemented via the
super-operator:
\be
\UB(z) = \Cal R e^{-\!\oint_{\! _z}\! \!  dw\,\Hbir(w)}.
\label{defUb}
\ee
Short-distance divergences in the expansion are exactly canceled by the regularization (point splitting) inherited from integrability, even out-of-equilibrium. This proves formula (\ref{defOeff}),
an expression that is finite order by order, and that straightforwardly generalizes to arbitrary super Fermi liquids.
For incoming fields, we find that $\UB\cdot\psiir_a(x<0)=\psiir_a(x<0)$ as expected from causality.

%%%%%%%%%%%%%%%%%%%%%%%%%%%%%%%%%%%%%%%%%%%%
\paragprl{(iii) Predictions}
\begin{figure}[h]
\includegraphics[width=0.45\textwidth]{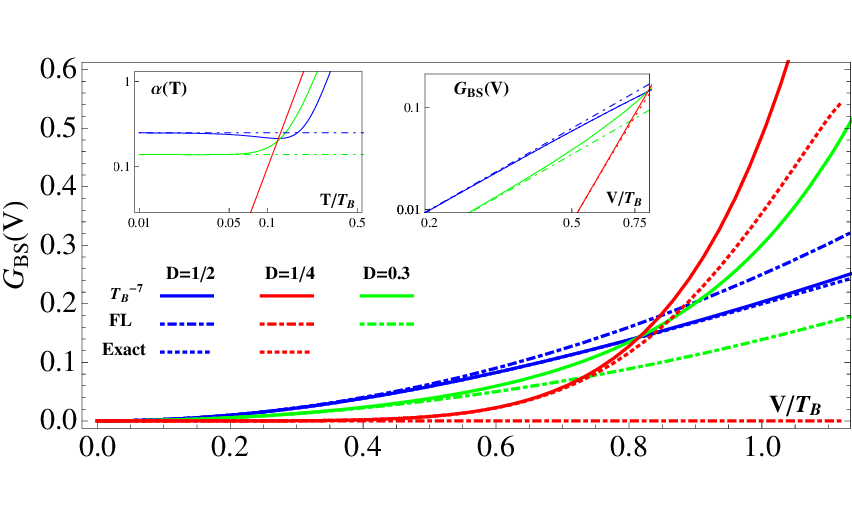}
\caption{[Color online] Differential conductance $\gbs(V)$ at $T\sh{=}0,\theta\sh{=}\frac{\pi}{2}$ in $\frac{e^2}{h}$ units. We also plot the exact results at $D=\frac{1}{2}$ (RLM) and $D\sh{=}\frac{1}{4}$ (SD-IRLM). Dotted lines are  FL results (up to $\TB^{-2}$). The inset shows the temperature dependence of $\alpha(T)$, the first non-linearity in 
$\frac{\gbs}{G_0}=\alpha(T)\frac{V^2}{\TB^2}+\Cal O(V^4)$.}
\label{IBSV}
\end{figure}
We calculate the effective symmetrized current operator $\hat I^{\eff}=\frac{\hat I_1^{\eff}-\hat I_2^{\eff}}{2}$ in the  IRLM by evaluating (\ref{defOeff}) \emph{at the operator level}  perturbatively in $\TB^{-1}$ at unprecedented
 large order  -- to do so,  we turn the problem into a purely algebraic one \cite{supplement} and solve it on a computer: this allows us to reach order $\TB^{-7}$ by  resuming $\sim 1500$ diagrams.

We first consider a DC bias $\mu_{1/2}=\pm V/2$  and compute  the back-scattered current $\ibs(V,T)=I_0^\ir-\langle\hat I^{\eff}\rangle_{\Hfreeir} = I_0^{\ir}\sum_{ij\geq0}a_{ij} V^i T^j/\TB^{i+j}$, where the coefficients  $a_{ij}(D)$ are pure numbers which we obtain exactly.
Our expressions match  the exact results available in the RLM and SD-IRLM\cite{irlmCurrent} limits, as well as  finite $T$ corrections in the linear regime \cite{boulat08}. The universal ratios 
$\frac{a_{04}}{a_{02}^2}
=\frac{3-12D-16D^2+g_4 12D(4D-1)(D-4)}{5(4D-1)}$ and
$\frac{a_{40}}{a_{20}^2}=\frac{3(8D^2-4D+1)+g_4 72D(D^2-3D+1)}{5(4D-1)}$
match with results  obtained at first order in $U$  at $V=0$\cite{doyon07} and $T=0$\cite{medenPrivate}.

Interestingly, we can prove \emph{non-perturbatively}\cite{supplement}, i.e. at any order in $\TB^{-1}$, that the back scattered current vanishes at imaginary voltage $V=\pm 2i\pi T$, i.e. the quantity  $\frac{\ibs(V,T)}{V^2+4\pi^2T^2}$ is an analytical function.

The resulting back-scattered non-linear conductance $\gbs(V)=\frac{\partial \ibs}{\partial V}$ is shown in Fig.\ref{IBSV}. We see the excellent agreement (up to $0.7 \,\TB$)  with  the exact expressions available in the 
RLM and SD-IRLM limits: it improves on the FL results (drastically when approaching the SD-IRLM).
\begin{figure}[h]
\includegraphics[width=0.235\textwidth]{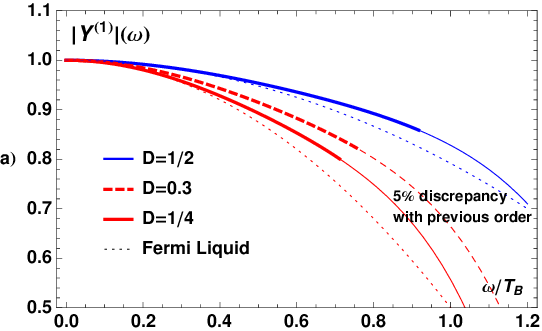} \;\includegraphics[width=0.235\textwidth]{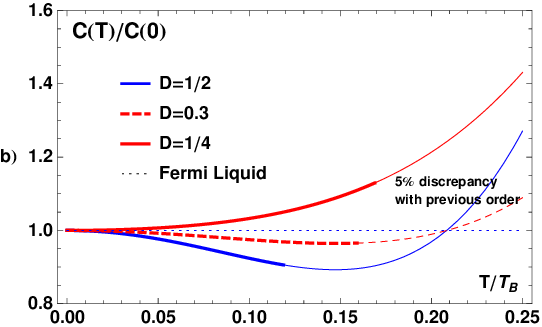}
\;
\includegraphics[width=0.235\textwidth]{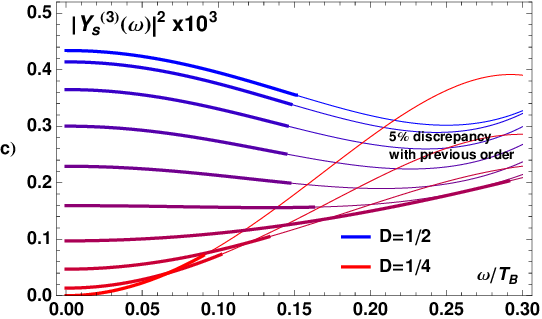}\;
\includegraphics[width=0.235\textwidth]{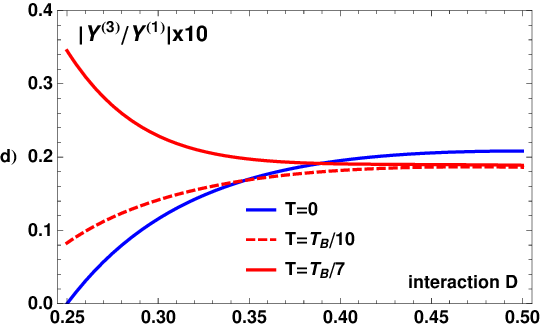}
\caption{[Color online] ($\theta\sh{=}\frac{\pi}{2}$) (a)  Modulus of the admittance $|\underline{Y}^{(1)}(\omega)|$ for an AC bias (in $e^2/h$ units).  The curves are dashed when the last two orders of our expansion differ by more than $5\%$. 
(b) Capacitance $C(\omega=0,T)$, normalized by the FL value $C(0,0)$.
(c) Amplitude of the $3\omega$  harmonics $\underline{Y}^{(3)}$ as a function of frequency ($T\sh{=}0$)
(d)  Influence of  temperature on the efficiency of harmonics production.}
\label{admittance}
\end{figure}

We then consider an AC bias $\mu_{1/2}(t)=\pm V/2\cos\omega t$. In the small voltage regime, the current can be written in terms of the admittance $\underline{Y}^{(1)}(\omega)=(G+i \omega C)(\omega)$ ($C$ is the capacitance):  $I(t)\sh{=}V\,\Re (\underline{Y}^{(1)} e^{i\omega t})$, and non-linear effects now lie in the $\omega$-dependence. 
We show in Fig.\ref{admittance} the amplitude $|\underline{Y}^{(1)}|$:
it is maximal in the RLM limit. Beyond the $T,\omega\to0$ limit $C(0,0) = \frac{e^2\cos2\theta}{4 D \TB h}$ predicted by FL theory, we find a rich dependence on $\omega$ and $T$, in particular $C(\omega\sh{=}0)$ (Fig. \ref{admittance}b) is significantly affected by temperature in a way that depends on the interactions.

We are also able to compute higher harmonics of the current, defining higher order admittances 
$\underline{Y}^{(n)}(\omega)= \lim_{V\to 0} \frac{\omega \TB^{n-1}}{\pi V^n} \int_{-\frac{\pi}{\omega}}^{\frac{\pi}{\omega}} I(t) e^{-i n \omega t} dt$.
In the particle-hole symmetric case $\epsilon_d=0$, 
the first non linear contribution is the third harmonic at $3\omega$. In Fig.\ref{admittance}c we  show how frequency affects the amplitude $|\underline Y^{(3)}|$. 
The FL predicts a constant $|\underline Y^{(3)}_{FL}|=\frac{(4D-1)\sin^2\theta}{192D^2}$ whereas our approach reveals that beyond a critical value of the interactions $D<D_c=0.3039...$, increasing frequency enhances the third harmonic $|\underline Y^{(3)}|(\omega)$. We also predict a strong enhancement of $|\underline Y^{(3)}/\underline Y^{(1)}|(0)$ at finite $T$ (see Fig.\ref{admittance}d).
Finally, breaking particle-hole symmetry with a grid potential $\epsilon_d d^\dagger d$, $2\omega$ harmonics are generated, and the lowest contribution to the admittance reads $\underline Y^{(2)}=-ig_4 \frac{3\cos\theta}{64 D^2}\frac{\epsilon_d\omega}{\TB^2}$ :
it does \emph{not} depend on the first process $\mathcal O_2$, and is thus completely missed by the FL approach.

%%%%%%%%%%%%%%%%%%%%%%%%%%%%%%%%%%%%%%%%%%%%
\paragprl{Conclusions}
We have shown that out-of-equilibrium  
QIM at their SC fixed point can be described by a simple deformation of the underlying conformal field theory describing the equilibrium SC fixed point, in the same spirit as finite temperature can be obtained as a mapping from the plane to the cylinder geometry.
Then, in integrable 
QIM that only allow for integer charge transfer (super Fermi liquids), we have shown that we can build a super-operator  mapping the out-of-equilibrium interacting theory onto a free one at equilibrium, yielding 
controlled, exact analytical results in the whole SC regime for the expansion of the universal scaling function of any physical quantity.
We have applied this method  to the IRLM, revealing  non-linear effects that are not captured by Fermi liquid theory.  
Our approach furnishes a mean to investigate in a controlled way many other out-of-equilibrium properties of super Fermi liquids, like (finite frequency) noise, full counting statistics, or non-linear thermal transport. Our approach could also be married with numerical techniques like diagrammatic Monte Carlo, by e.g. appropriately sampling the set of Feynman diagrams contributing to physical quantities.
It also raises the question whether it can be generalized to   non (super) Fermi liquid integrable 
QIM.

\breakprl
 
Acknowledgements:  E.B. would like to thank S. Florens, O. Parcollet, I. Paul, T. Popov and H. Saleur for stimulating  discussions and encouragements. L.F.  acknowledges support from the French DGA.

%%%%%%%%%%%%%%%%%%%%%%%%%%%%%%%%%%%%%%%%%%%%%%%%%%%%%%%%%

\bibliographystyle{unsrt}

\clearpage
\begin{widetext}
\begin{center}

{\bf\large Supplementary material for :\\ \vspace{1cm}
Out-of-equilibrium properties and non-linear effects for interacting quantum impurity systems in their strong coupling regime}

\vspace*{0.5cm}

L. Freton, and E. Boulat\\
\emph{Laboratoire Mat\'eriaux et Ph\'enom\`enes Quantiques,\\
Universit\'e Paris  Diderot, CNRS UMR 7162, 75013 Paris, France}

\vspace*{0.5cm}
\end{center}
\end{widetext}

\newcommand{\Uferm}{U_{\text{F}} }
\newcommand{\Ubos}{U_{\text{B}}}
\newcommand{\Ulatt}{U_{\text{latt}}}

\section{Weak/ and Strong/coupling theories}

We start by presenting the two dual descriptions of the IRLM, either in the weak coupling regime (this is the standard "microscopic theory"), or in the strong coupling regime.

\subsection{Weak coupling Hamiltonian}

The IRLM is defined by the microscopic theory:
\bea
H&=&\sum_{a} H_0[\psi_a]+\HB,
\label{defHirlmsupp}\\
H_0[\psi]&=&-i\vf \!\!\int_{-\infty}^\infty \!\!\!dx\,\psi^\dagger\p_x\psi^{\vphantom\dagger}
\label{defH0irlmsupp} 
\\
\HB&=&\gamma_a\psi^\dagger_a(0)d +\hc + U \!:\!\psi^\dagger_a\psi_a^{\vphantom\dagger}\!\!:\!(0)\big(d^\dagger d -\textstyle{\frac{1}{2}}\big)\nonumber\\
&&+ \epsilon_d \,d^\dagger d,
\label{defHBRILMsupp} 
\eea
%we introduce the fermions $\psi_\pm=\frac{1}{\sqrt{\gamma_1^2+\gamma_2^2}}\left(\gamma_{1(2)}\psi_1\pm\gamma_{2(1)}\psi_2\right)$
It can be mapped onto the anisotropic Kondo model by introducing two bosons $\phi_{\pm}$  \cite{boulat08supp} in terms of which the Hamiltonian reads:
\bea
H&=&\sum_{\pm} H_0[\phi_\pm]+\HB\nonumber\\
H_0[\phi]&=&\vf\int dx\;(\p_x \phi)^2\nonumber\\
\HB&=&\gamma\,e^{i\beta\phi_+}(0)S^+ +\hc+\epsilon_d\,S^z
\label{defHBK} 
\eea
with the spin operators $S^z=d^\dagger d-\frac 1 2 $, $S^+=\eta d^\dagger$, $S^-=(S^+)^\dagger$ and $\eta$ a local Majorana fermion, $\eta^\dagger=\eta^{-1}=\eta$. The gradients of the bosons define charges $Q_{\pm}=\sqrt{2\pi}\,\p_x\phi_\pm$ that are linearly related 
to fermionic bilinears. The mapping from (\ref{defHBRILMsupp}) to (\ref{defHBK}) involves a unitary operation that allows to absorb the coulombic repulsion term $U$ in a redefinition of the tunneling operator. The latter acquires an anomalous scaling dimension $D(U)=\frac{\beta^2}{8\pi}$ that will in general depend on the way one regularizes the $\delta$ function in front of the $U$ term in (\ref{defHBRILMsupp}). In the half-filled lattice regularized IRLM, used e.g. in numerical simulations, with Hamiltonian $H=-t\sum_{j\neq 0}c^\dagger_{j}c\vph_{j+1}+\hc+ \big[c^\dagger_0(\gamma_1 c\vph_1+\gamma_2 c\vph_{-1})+\hc\big] +U_{\ind{\textsc{latt}}}\big(c^\dagger_0c\vph_0-\frac{1}{2}\big)\big(c^\dagger_1c\vph_1+c^\dagger_{-1}c\vph_{-1}-1\big)$, one has 
\be
D(\Ulatt)=\frac{1}{4}+ \big(\frac{\arctan\frac{1}{2}(\frac{(\Ulatt)}{2t}-\frac{2t}{(\Ulatt)})}{\pi}\big)^2.
\ee
The relationship between $\Ulatt$ and the  parameter $U$ of the fermionic continuous theory (\ref{defHirlmsupp}) being $U=4\Ulatt a_0$, with $a_0 $ the lattice spacing.
$D(U)$ is sometimes given \cite{irlmCurrent} with an $U=\Ubos$ defined in a bosonic version of (\ref{defHBRILMsupp}), the relation in the cut-off scheme associated with abelian bosonization is $\Ubos /4\vf = \arctan(U /4\vf)$, see e.g. Ref.\cite{schillerandrei}.

\subsection{Strong coupling Hamiltonian}

Under the renormalization group, the coupling $\gamma$ flows to strong coupling at low energy $E\lesssim \TB$.  
At the strong coupling fixed point $E=0$ (or equivalently $\TB=\infty$), the theory is simply described by 
\be
H_0^\ir=\sum_{\pm}H_0[\phi_\pm]
\ee
where the boson $\phi_+$ has a phase shift at the origin, $\phi_+(0^+)=\phi_+(0^-)+\frac\beta 4$ (the boson $\phi_-$ is unaffected and obeys trivial boundary conditions $\phi_-(0^+)=\phi_-(0^-)$). Building on some remarkable results about the description of integrable boundary theories in terms of "boundary state" \cite{ghoshal},
it  could been shown \cite{lesage99a,lesage99b} that a  dual description of the anisotropic Kondo Hamiltonian (\ref{defHBK})  can be derived:
\be
\HB^\ir=\sum_{n=1}^\infty\frac{g_{2n}\vph}{ \TB^{2n-1}}\Cal O_{2n}(x=0)
\label{HBIRsupp}
\ee
 where the couplings read\cite{lesage99b}:
\be
g_{2n}\vph = \frac{\left(\frac{D}{\pi}\right)^{n-1}}{(n-\frac{1}{2})\,n!}\;
\frac{\Gamma\left(\frac{2n-1}{2(1-D)} \right)\Gamma\left(\frac{D}{2(1-D)} \right)^{2n-1} }{\Gamma\left(\frac{1}{2(1-D)} \right)^{2n-1} \Gamma\left(\frac{(2n-1)D}{2(1-D)} \right)}
\label{couplings}
\ee
whereas the operators $\Cal O_{2n}$ are the conserved quantities of the sine Gordon model and can be built out of an elementary field, a modified stress energy tensor $\Cal O_2$. The first members of the series explicitly read \cite{eguchiyang89,lesage99a}
\bea
\Cal O_2 &=& (Q_+^2)
-\alpha_0\p Q_+ \;,\quad \alpha_0 = \frac{1-D}{\sqrt{D}}\;,\quad c=1-6\alpha_0^2 
\vphantom{\frac{c}{6}}
 \nonumber\\
 \Cal O_4 &=& (\Cal O_2\Cal O_2)\vphantom{\frac{c}{6}}
  \nonumber\\
  \Cal O_6 &=& (\Cal O_2\Cal O_4) +\frac{c+2}{12}\;(\Cal O_2\p^2\Cal O_2)
%  \nonumber\\
 %   \Cal O_8 &=&  (\Cal O_2(\Cal O_2\Cal O_4)) + \frac{c+8}{6}\;(\Cal O_2(\Cal O_2\p^2\Cal O_2))
  %\nonumber\\    
   % &&+\frac{c^2+4c-101}{180}(\Cal O_2\p^4\Cal O_2)
 \eea
 The modified stress energy tensor is built out of the U(1) charge $Q_+=M_{ab}:\Psi^\dagger_a\Psi\vph_b:$, with 
 \be
 M=\frac{\pi}{\sqrt{4D}}\left(\!\!\begin{array}{cc}
  \sqrt{4D-1}+\cos\theta& \sin\theta\\
\sin\theta  &  \sqrt{4D-1}-\cos\theta
 \end{array}
 \!\!\right). 
 \ee

\section{Current at the SC fixed point}

Here we describe the steps needed to obtain the electrical current at the strong coupling fixed point (SC-FP), i.e. when all physical scales $E=V,T,\epsilon_d,\omega...\ll \TB$, or equivalently when $\TB=
\infty$.  The electrical current operator $j_a(x)$ on wire $a$ at a point $x<0$, is obtained through the continuity equation for the charge density $n_a(x)=:\!\psi_a^\dagger\psi_a\vph\!\!:(x)\;+:\!\psi_a^\dagger\psi_a\!\!:(-x)$, and reads $j_a(x)=:\!\psi_a^\dagger\psi_a\vph\!\!:(x)\;-:\!\psi_a^\dagger\psi_a\!\!:(-x)$. The current is then expressed in term of the "transparent fields" $\Psi_a$ that do not feel the impurity (i.e. that have trivial boundary condition at the SC fixed point). The boundary conditions obeyed by the original fields at the SC FP are encoded in the relationship between original and transparent fields. They are conveniently expressed in terms of the densities  $Q_{ab}\equiv \density{\psi}{a}{b}$ and $\mathcal Q_{ab}\equiv \density{\Psi}{a}{b}$. While incoming densities of transparent and original fields coincide, $\mathcal Q_{ab}(x<0)=Q_{ab}(x)$ (this traduces causality, i.e. the fact that at the  left of the impurity, right moving fields are not influenced by the impurity), outgoing densities do obey:
\bea
\mathcal Q_{ab}(x>0)&=&\text{B}(\theta)_{ab}^{cd}\,Q_{cd}(x),
\\
 \big[\text{B}(\theta)\big]_{ab}^{cd}&=&\cos^2\theta\;\sigma^3\otimes\tau^3 + \cos^2\theta\;\sigma^1\otimes\tau^1\nonumber\\
 &&+\sin\theta\cos\theta\;\left(\sigma^1\otimes\tau^3+\sigma^3\otimes\tau^1\right)
 \nonumber
 \eea
The $\text{B}$ matrix encoding the SC boundary conditions involves Pauli matrices $\sigma$ ($\tau$ respectively) acting on the pair of indices ($a,c$) (on the pair ($b,d$) respectively). Taking the limit $x\to 0^-$ yields the current through the impurity  in terms of transparent fields, 
\be
\hat I_a=j_a(0^-)=(\delta_a^c\delta_a^d-\text{B}^{cd}_{aa}(\theta)):\!\!\psiir_c^\dagger\psiir_d\vph\!:(0,t).
\ee

Then, to force the system out-of-equilibrum at the SC fixed point, we  apply $\Ugauge_{\NEQ}$ to obtain the effective current operator $\hat I_a^{\eff}=\Ugauge_{\NEQ}\cdot \hat I_a$, whose equilibrium average value gives the non-equilibrium average value of the original current operator. The transformation $\Ugauge_\NEQ$ mixes descendent fields amongst each other, and we find that the effective current acquires a part proportional to the identity which will give rise to the SC, non equilibrium expectation value
\begin{equation}
\hat I_a^{\eff}(t)
\sh{=}\sum_{b,c} \Lambda_{a}^{bc}(t)\;:\!\psiir_b^\dagger\psiir_c\vph\!\!:(0,t)\sh{+}\frac{\sin^2\theta}{2\pi}(\mu_1( t)-\mu_2(t))\;\mathbb{I}
\label{Iasc}
\end{equation}
with explicitly 
\begin{equation}
\Lambda_a^{bc}(t)=\frac{(-)^{a+1} }{2}
\begin{pmatrix}
2 \sin^2\theta &\sin2\theta \,e^{(\MU_1-\MU_2)(it)}\\
\sin2\theta \, e^{-(\MU_1-\MU_2)(it)}& 2\sin^2\theta 
 \end{pmatrix}_{bc}.
\end{equation}

Taking the expectation value w.r.t the thermal density matrix $e^{-\beta H_0[\Psi]}$, we recover the linear regime:
\begin{equation}
I_a^\ir \equiv \langle \hat I_a^{\eff} \rangle_{H_0^\ir}=\frac{\sin^2\theta}{2\pi}(\mu_1 (t)-\mu_2(t)).
\end{equation}
The current is linear in the bias, irrespective of the temperature (recall that $\TB=\infty$ here: the system has conformal symmetry and the finite temperature simply obtained via the usual mapping to the cylinder geometry, that does not affect the v.e.v. of the current (\ref{Iasc})).

\section{Around the SC fixed point}

\subsection{Computation details}

Departing from the SC fixed point amounts to setting $\TB<\infty$ ; the effective current operator now acquires contributions from back-scattering processes: it will become dressed by a cloud of particle-hole excitations. We write the effective current operator 
\bea
\hat I_a^{\eff}&=&\Ugauge_{\NEQ}\cdot \hat I_a^{\eff,0}, \nonumber\\
\hat I_a^{\eff,0} &=& \UB \cdot \hat I_a,
\eea
 clearly separating $(i)$ on the one hand the effect of the interactions that dress the current operator by a cloud of particle-hole pairs yielding $\hat I_a^{\eff,0}$ (this dressing is done \emph{at the fixed point}), and $(ii)$ on the other hand the boundary conditions imposed on incoming fields (at the far left of the impurity), i.e. incoming density matrices representing free Fermi seas biased by time dependent potentials encoded in an incoming Hershfield operator $Y^{\mbox{\scriptsize{in}}}=\int_{-\infty}^\infty dx\;\mu_a \density{\Psi}{a}{a}$.

We expand the current 
\be
\hat I_a^{\eff,0}=\sum_{n\geq 0} \TB^{-n}\;\hat I_a^{(n)}
\ee
 by expanding Eq. (1) in the main text, i.e. by taking systematically into account  the perturbation $\HB^\ir$ (\ref{HBIRsupp}). 

%%%

 The perturbative expansion can be explicitly carried out  at the level of operators. At  each order  $\hat I_a^{(n)}$  bears contributions from all patterns $\hat  {\cal P}^{\bold n}$ of insertions of perturbing operators $\Cal O_{2n_i}$, labelled by  $\bold{n}=\{n_1,n_2,...,n_k\}$ with the constraint 
$\sum_i (2n_i-1)=n$.  Those operators are inserted along Keldysh contours. Using a super operator formulation \cite{harbola08} of the Keldysh expansion, we write the effective current operator as:
\begin{widetext}
\bea
\hat I_a^{(n)}(t)&=&
\sum'_{\bold{n}}  C_{\bold{n}}\prod_{i=1}^k g_{2n_i} \sum_{bc}\Lambda_a^{bc}(t)\;\hat{\cal P}_{bc}^{\,\bold{n}}(t)
\nonumber\\
\hat{\cal P}_{ab}^{\,\bold{n}}(t)&=&\int_{-\infty}^t dt_1\int_{-\infty}^{t_1} dt_2... \int_{-\infty}^{t_{k-1}} dt_k [\Cal O_{2n_1}(t_1),[\Cal O_{2n_2}(t_2),...[\Cal O_{2n_k}(t_k),:\!\!\psiir_a^\dagger\psiir_b\vph\!\!:(t)]...]]\nonumber\\
&=&\Cal O_{2n_1}\star...\star\Cal O_{2n_k}\star:\!\!\psiir_a^\dagger\psiir_b\vph\!\!:
\label{defFn}
\eea 
\end{widetext}
Here the symbol $\sum'_{\bold{n}}$ means that it is restricted to $\bold{n}$'s satisfying the constraint $\sum_i n_i=\frac{n+k}{2}$, and  $C_{\bold{n}}$ is a combinatoric factor.

The binary operation $"\star"$ is in general a complicated, non-associative, non-commutative and non-local operation in the space of local operators of the CFT describing the SC-FP. In a super Fermi liquid, the perturbing operators $\Cal O$'s are \emph{integer fields} (specifically and technically, Kac Moody descendants of the identity operator \cite{difrancesco}, specifically and physically, fields built out of the fermion fields $\Psi_a\vph$ and $\Psi_a^\dagger$). It results that nested commutators of  $\HB^\ir$ with $\density{\Psi}{a}{b}$ can be represented as contour integrals (here we use the crucial property that the operator product expansion $\HB^\ir(z)A(w)=\sum_k\frac{\{\HB^\ir A\}_{_k}(w)}{(z-w)^k}$ only has integer exponents $k$ (i.e. $\HB^\ir$ and $A$ are mutually local operators) if $A$ is a fermonic bilinear  operator $\density{\Psi}{a}{b}$, and this property holds recursively, i.e. replacing $A$ by $\{\HB^\ir A\}_{_k}$ etc...).

\begin{figure}[h]
\[
\sum_{n=0}^\infty (-i)^n  \int_{\Cal C_K}\vph \big(\prod_{j=1}^n dt_j\big) \;\;\;
\raisebox{-0.55\height}{\includegraphics[width=0.17\textwidth]{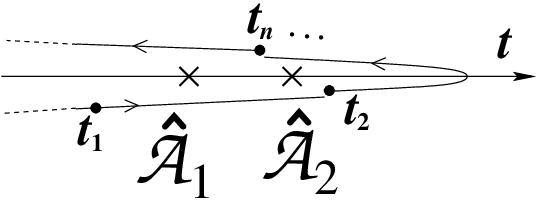}} 
\]
\[
= \; \sum_n \frac{(-i)^n}{n!}\big(\prod_{j=1}^n \oint dt_j \big)\;
\raisebox{-0.55\height}{\includegraphics[width=0.11\textwidth]{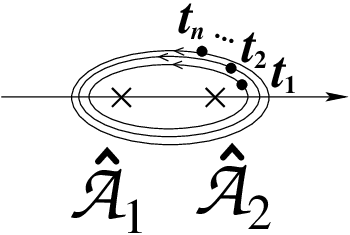}} \;\;\;
= \;\;
\raisebox{-0.77\height}{\includegraphics[width=0.11\textwidth]{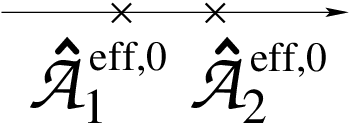}} 
\]
\caption{
Effective operators in super Fermi liquids. The symbols $t_i$ indicate insertions of the operator $\Hbir(t_i)$. The  Keldysh expansion of a $N$-point correlator (here $N=2$) on top is formally resumed by introducing effective, dressed operators defined in (\ref{defAeff}), on the bottom. In the intermediate step, this correlator is written an exponential of nested (radially ordered) contour integrals.
}
\label{sketchKeldysh}
\end{figure}

In a super Fermi liquid, the operation "$\star$" on the space of local operators of the theory thus takes a simple and compact form:  
\be
\Cal O \star\Phi(z)\equiv\oint_{\!_z} dw \;\Cal O(w)\Phi(z)=2i\pi\;\{\Cal O\;\Phi\}_1\vph(z).
\label{defStar}
\ee
so that the whole Keldysh expansion can be formally resumed (see Fig.\ref{sketchKeldysh}) by defining the effective operators:
\bea
\hat{\cal A}^{\eff,0}(z)=\UB(z)\cdot \hat{\cal A}(z)=\Cal R e^{-\!\oint_{\! _z}\! \!  dw\,\Hbir(w)}\hat{\cal A}(z)
\nonumber\\
%=\sum_{k=0}^\infty\frac{(-2i\pi)^k}{k!}  \{\Hbir\; \{\Hbir\; \ldots\{\Hbir\;  \hat{\cal A} \}_1\vph\ldots  \}_1\vph \}_1\vph 
=\sum_{n=0}^\infty\frac{(-1)^n}{n!}  
\Big( 
\underbrace{\Hbir\star\Hbir\star \ldots\star \Hbir}_{n\vph} \; \star\;  \hat{\cal A}
\Big)(z)
\label{defAeff}
\eea

Being \emph{commuting} conserved quantities, the operator product expansion (OPE) \cite{difrancesco} between two $\Cal O$'s has the property that the (operator-)coefficient of the term $(z-w)^{-1}$ is a total derivative, resulting in the operation $\star$ being associative and commutative ; this simplifies the combinatorics of the expansion. 

The determination of $\hat I_a^{\eff,0}$  becomes a purely algebraic problem:
at order $\TB^{-n}$, $\hat I_a^{(n)}$ lives in the space $\Cal E_{n+1}$ of  Kac Moody descendent fields \cite{difrancesco} of the identity operator
with conformal weights $\leq n+1$, and all one needs to known is how to actually evaluate (\ref{defAeff}), or equivalently, using (\ref{defStar}), the coefficient in $(z-w)^{-1}$ in the OPE $\Cal O(z)\Phi(w)=\sum_{n}\frac{\{\Cal O\,\Phi\}_{n}\vph(w)}{(z-w)^n}$ for an arbitrary field $\Phi$ in $\Cal E_n$. This can be done recursively by using the elementary OPE $\Psi_a^\dagger(z)\Psi\vph_b(w)=\frac{\delta_{ab}}{2\pi(z-w)}$ + \mbox{regular}. 
The space $\Cal E_{n}$ is spanned by elementary operators $\hat{\cal F}_{\bold{p},\bold{\bar p},\bold{q},\bold{\bar q}}$ that simply correspond to all physical processes transferring electrons from one wire to another at the impurity site, with the simultaneous emission of an arbitrary number of particle-hole pairs:
\bea
&\hat{\cal F}_{\bold{p},\bold{\bar p},\bold{q},\bold{\bar q}}= \hat{\cal F}_{\{p_1,...,p_{m_1},\bar p_1,...,\bar p_{\bar m_1},q_1,...,q_{m_2},\bar q_1,...,\bar q_{\bar m_2}\}}
\;\;=\nonumber\\
&(\p^{p_1}\Psi_1^\dagger( ... \p^{p_{m_1}}\Psi_1^\dagger(\p^{\bar p_1}\Psi_1\vph(...\p^{\bar p_{\bar m_1}}\Psi_1\vph
(\p^{q_1}\Psi_2^\dagger( ...
\nonumber\\
&... \p^{q_{m_2}}\Psi_2^\dagger(\p^{\bar q_1}\Psi_2\vph(...\p^{\bar q_{\bar m_2}}\Psi_2\vph
))...)))
\label{defF}
\eea
 involving an arbitrary even number $2p=m_1+\bar m_1+m_2+\bar m_2$  of fermions, with $p=m_1+m_2=\bar m_1+\bar m_2$.  Fermi statistics allows one to choose ordered sets 
 $p_{i}>p_{i+1}$, 
and similarly for the $q_i$'s, the $\bar p_i$'s, and the $\bar q_i$'s. The nested parenthesis indicate the normal order product of two local operators $A(z)$ and $B(w)$, $(AB)(w)\equiv\frac{1}{2 i\pi}\oint_w\frac{dz}{z-w}\,A(z)\,B(w)$.
  Each  process (\ref{defF}) is thus an elementary  Feynman diagram, and one has to compute two things: $(i)$  the weight of each elementary diagram (\ref{defF}) in the expansion of $\hat I_a^{\eff,0}$ and $(ii)$ the out-of-equilibrium thermal expectation value of  each elementary diagram $\Ugauge_{\NEQ}\cdot \hat{\cal F}_{\bold{p},\bold{\bar p},\bold{q},\bold{\bar q}}$.  

The dimension of $\Cal E_{n}$ grows as $\sim n^{-\frac{1}{2}}e^{2\pi\sqrt{\frac{n}{3}}}$ \cite{hardyramanujan},  so actual calculations quickly become  extremely tedious, since they e.g. require the computation of the full OPE for two arbitrary   operators in $\Cal E_n$.    We resort to a code, a calculator in $\Cal E_n$, to evaluate the $\simeq 1500$ diagrams necessary to reach order $n=6$. 
Elementary calculations in $\Cal E_n$ are heavily recursive stemming from the fact that basic operations (like the (normal-ordered) product of two local operators,  or more generally the OPE between two local operators) are neither commutative nor associative.

\subsection{Explicit dressed current operator}

Here we give explicitly, for illustration,  the expression of the dressed (back-scattered) current  operator,
$\hat I_\bs^{\eff,0}=(\id -\UB)\cdot \hat I$
(we consider the symmetrized current $\hat I=\frac{\hat I_1-\hat I_2}{2}$),
on the basis of the set of diagrams (\ref{defF}). Since expressions quickly become cumbersome, we show only the expansion
$\hat I_\bs^{\eff,0}=\sum_{n \geq 0} \TB^{-n} \hat I_\bs^{(n)}$ up to $\TB^{-2}$ and we denote $C=\cos\theta$
\begin{widetext}
\begin{align}
\hat I_\bs^{(0)}=0
\label{currop0}
\end{align}
%\begin{align}
%\hat I_\bs^{(0)}=
%\sqrt{1-C^2} C (\left(\psiir_1^\dagger\psiir_2^{\vphantom{\dagger}}\right)-\left(\psiir_1^{\vphantom{\dagger}}\psiir_2^\dagger\right))
%+(C^2-\frac{1}{2}) (\left(\psiir_1^\dagger\psiir_1^{\vphantom{\dagger}}\right)-\left(\psiir_2^\dagger\psiir_2^{\vphantom{\dagger}}\right))
%\label{currop0}
%\end{align}
%
\begin{align}
\hat I_\bs^{(1)}
=&
\frac{i}{4 D}
\left(1-2 C^2-C \sqrt{4 D-1}\right) 
\left[
\left(\partial\psiir_1^\dagger\psiir_1^{\vphantom{\dagger}}\right)+\left(\psiir_1^\dagger\partial\psiir_1^{\vphantom{\dagger}}\right)
\right]\nonumber\\
&-
\frac{i}{4 D}
\left(1-2 C^2+C \sqrt{4 D-1}\right) 
\left[
\left(\partial\psiir_2^\dagger\psiir_2^{\vphantom{\dagger}}\right)+\left(\psiir_2^\dagger\partial\psiir_2^{\vphantom{\dagger}}\right)
\right]
\nonumber\\
&+
\frac{i}{4 D}
\sqrt{1-C^2} \left(2 C+\sqrt{4 D-1}\right) 
\left[
\left(\partial\psiir_1^{\vphantom{\dagger}}\psiir_2^\dagger\right)-\left(\partial\psiir_1^\dagger\psiir_2^{\vphantom{\dagger}}\right)
\right]
\nonumber\\
&+
\frac{i}{4 D}
\sqrt{1-C^2} \left(2 C-\sqrt{4 D-1}\right) 
\left[
\left(\psiir_1^{\vphantom{\dagger}}\partial\psiir_2^\dagger\right)-\left(\psiir_1^\dagger\partial\psiir_2^{\vphantom{\dagger}}\right)
\right]
\label{currop1}
\end{align}

\begin{align}
\hat I_\bs^{(2)}
=&
\frac{1}{4 D}
\left(1-2 C^2-C \sqrt{4 D-1}\right) 
\left[
\left(\partial^2\psiir_1^\dagger\psiir_1^{\vphantom{\dagger}}\right)+\left(\psiir_1^\dagger\partial^2\psiir_1^{\vphantom{\dagger}}\right)
\right]
\nonumber\\
&-\frac{1}{4 D}
\left(1-2 C^2+C \sqrt{4 D-1}\right)
\left[
\left(\partial^2\psiir_2^\dagger\psiir_2^{\vphantom{\dagger}}\right)+\left(\psiir_2^\dagger\partial^2\psiir_2^{\vphantom{\dagger}}\right)
\right]\nonumber\\
&+
\frac{1}{4 D^2}
C \sqrt{1-C^2} \left[
\left(\partial\psiir_1^{\vphantom{\dagger}}\partial\psiir_2^\dagger\right)-\left(\partial\psiir_1^\dagger\partial\psiir_2^{\vphantom{\dagger}}\right)
\right]
\nonumber\\
&+
\frac{1}{16 D^2}
\sqrt{1-C^2} \left(8 C D+(2 D+1) \sqrt{4 D-1}\right) 
\left[
\left(\partial^2\psiir_1^{\vphantom{\dagger}}\psiir_2^\dagger\right)-\left(\partial^2\psiir_1^\dagger\psiir_2^{\vphantom{\dagger}}\right)
\right]
\nonumber\\
&+
\frac{1}{16 D^2}
\sqrt{1-C^2} \left(8 C D-(2 D+1) \sqrt{4 D-1}\right) \left[\left(\psiir_1^{\vphantom{\dagger}}\partial^2\psiir_2^\dagger\right)-\left(\psiir_1^\dagger\partial^2\psiir_2^{\vphantom{\dagger}}\right)
\right]
\nonumber\\
&+
\frac{1}{8 D^2}
\sqrt{1-C^2} (1-2 D) \sqrt{4 D-1}
\big[
\left(\psiir_1^\dagger\left(\psiir_2^\dagger\left(\partial\psiir_2^{\vphantom{\dagger}}\psiir_2^{\vphantom{\dagger}}\right)\right)\right)
\nonumber\\
&\hspace{1cm}-\left(\partial\psiir_1^\dagger\left(\psiir_1^\dagger\left(\psiir_1^{\vphantom{\dagger}}\psiir_2^{\vphantom{\dagger}}\right)\right)\right)
%\nonumber\\
%&\hspace{5cm}
+\left(\psiir_1^\dagger\left(\partial\psiir_1^{\vphantom{\dagger}}\left(\psiir_1^{\vphantom{\dagger}}\psiir_2^\dagger\right)\right)\right)
%\nonumber\\
%&\hspace{5cm}
-\left(\psiir_1^{\vphantom{\dagger}}\left(\partial\psiir_2^\dagger\left(\psiir_2^\dagger\psiir_2^{\vphantom{\dagger}}\right)\right)\right)
\big]
\nonumber\\
&+\frac{1}{4 D^2}
\left(C^2-2 C D \sqrt{4 D-1}+2 D-1\right)
\left(\partial\psiir_2^\dagger\partial\psiir_2^{\vphantom{\dagger}}\right) 
\nonumber\\
&-\frac{1}{4 D^2}
\left(C^2+2 C D \sqrt{4 D-1}+2 D-1\right)
\left(\partial\psiir_1^\dagger\partial\psiir_1^{\vphantom{\dagger}}\right) 
\label{currop2}
\end{align}

\end{widetext}

\subsection{Current expansion}

To evaluate the expectation value of the current in the non-equilibrium theory, one needs to act with the super-operator $\Ugauge_\NEQ$ on the dressed current (\ref{currop0}-\ref{currop2}) ; in doing so one generates a component on the identity operator, that directly yields a finite expectation value.
We give here the expression of the first terms of the backscattered current 
$\ibs=\big\langle \Ugauge_\NEQ \cdot \hat I_\bs^{\eff,0}\big\rangle_{\Hfreeir}$ for a static applied bias $V$, at $T\neq 0$ and in the resonant case $\epsilon_d=0$ 
\begin{widetext}
\be
\frac{\ibs}{I^{\ir}}
=\frac{X (V^2+4\pi^2  T^2)}{48\,D^2\,\TB^2}\Big[1+
\frac{3g_4(8\pi^2   T^2 X(X-15)+3V^2(X^2-10X+5))}{40D \TB^2} 
-\frac{ 8\pi^2   T^2(1+5X+X^2)+3V^2(X^2+1)}{40D^2\TB^2} \Big]+\mathcal O\big(\TB^{-6}\big)
\ee
\end{widetext}
with $X=4D-1$, $I^\ir = (I_1^\ir - I_2^\ir)/2$ and the coupling $g_4$ is given by Eq.(\ref{couplings}).
One notices that the quantity $V^2+4\pi^2 T^2$ factorizes in the expression of $\ibs(V,T)$. This property can be shown to hold at arbitrary order in perturbation theory, and can be seen to be a direct consequence of the respective nature of finite voltage and finite temperature from the viewpoint of the deformations of the CFT (SC-FP) they are represented by. Indeed, to evaluate the contribution to the average value of a diagram $\hat{\cal F}_{\bold{p},\bold{\bar p},\bold{q},\bold{\bar q}}$, where the average in taken in the finite temperature, non-equilibrium SC theory, one needs to first absorb the voltage (one writes $z=i(t-x)$) by the change $\Psi_a\vph(z)\to e^{i(-)^a V(t-x)/2}\,\Psi_a\vph(z)$, and then the finite temperature effect by a mapping of the cylinder to the plane ($z\to\omega=e^{2i\pi T z}$), $\Psi_a\vph(z)\to \sqrt{2i\pi T}\;e^{i\pi T z}\;\Psi_a\vph(\omega)$. All together, when choosing $V=2i\pi T$ (when $V=-2i\pi T$, the role of $\Psi\vph_1$ and $\Psi\vph_2$ are exchanged)  the fermions are changed according to:
\bea
\Psi_1\vph(z)&\to& \sqrt{2i\pi T} \;\;\Psi_1\vph(\omega) \nonumber\\
\Psi_1^\dagger(z)&\to& \sqrt{2i\pi T} \;\;\Psi_1^\dagger(\omega)\;\omega \nonumber\\
\Psi_2\vph(z)&\to& \sqrt{2i\pi T} \;\;\Psi_2\vph(\omega) \;\omega \nonumber\\
\Psi_2^\dagger(z)&\to& \sqrt{2i\pi T} \;\;\Psi_2^\dagger(\omega) \nonumber
\eea
More generally, under the finite temperature and finite voltage deformation of the CFT, any local operator transforms 
in a sum of new operators with coefficients that are \emph{polynomials} in the variable $\omega$. In the IRLM, this results in the fact 
that the identity operator cannot appear in the back-scattered current under the action of the finite voltage 
and finite temperature deformations, so that $I_\bs(\pm 2 i T,T) = 0$ at any order of perturbation theory. 
%%%%%%%%%%%%%%%%%%%%%%%%%%%%%%%%%%%%%%%%%%%%%%%%%%%

\bibliographystyle{unsrt}

\end{document}